\documentclass[5p,twocolumn,times,number]{elsarticle}

\usepackage{graphicx}
\usepackage{amsmath}   
\usepackage{subfigure} 
\usepackage{footnote}

\newcommand{\sr}{~\ensuremath{^{90}\mathrm{Sr}} }

\begin{document}

\begin{frontmatter}

\title{ Radiation Tolerance of the LHCb Outer Tracker: in the Lab and in the Forward Region at the LHC \footnotemark[1]}   

\author[add1]{N.Tuning\corref{cor}}
\author[add2]{S.Bachmann}
\author[add1]{A.Pellegrino}
\ead{tuning@nikhef.nl}
\author[add2]{U.Uwer}
\author[add3]{D.Wiedner}

\cortext[cor]{Corresponding author}

\address[add1]{Nikhef, Amsterdam, The Netherlands}
\address[add2]{Physikalisches Institut, Heidelberg, Germany}
\address[add3]{Technische Universit\"at Dortmund, Germany}

\begin{abstract}
During the detector construction phase between 2004 and 2006,
it was discovered that
the LHCb Outer Tracker (OT) detector suffered from gain loss after
irradiation in the laboratory at moderate intensities. Under irradiation an
insulating layer was formed on the anode wire.  The aging was caused by
contamination of the counting gas due to outgassing of the glue used in
construction namely araldite AY103-1.  The gain loss was concentrated upstream
the gas flow, and at moderate irradiation intensity only.  The aging rate was
reduced by longterm flushing and by the addition of a few percent of O$_2$ to
the gas mixture. Furthermore, applying a large positive high voltage (beyond the
amplification regime) has shown to remove the insulating deposits without damaging the wire
surface.  This paper presents the history of the developments together with the characteristics and the culprit
of the aging phenomenon and the resulting detector performance in situ.
\end{abstract}


\end{frontmatter}

\footnotetext[1]{Presented at 3rd Int.Conf. on Detector Stability and Aging Phenomena in Gaseous Detectors, 6 Nov 2023.}

\section{The LHCb Outer Tracker}
The LHCb experiment has provided a wealth of  CP violation and rare $B$-decays measurement
fulfilling its promise and beyond, surpassing its original physics programme with world-leading 
results in the areas of heavy-ion physics and in the high-$p_T$ electroweak sector.
These phenomenal successes were possible despite the original worries regarding the 
radiation hardness of the Outer Tracker (OT) detector, a gaseous straw tube detector~\cite{LHCbOuterTrackerGroup:2013epe}
that covered an area of approximately 5x6~m$^2$ with 12 double layers of straw
tubes.

\begin{figure}[!h]
\begin{center}
    \begin{picture}(250,90)(0,0)
    \put( 0,20){\includegraphics[bb=0 0 1982 875,scale=0.08]{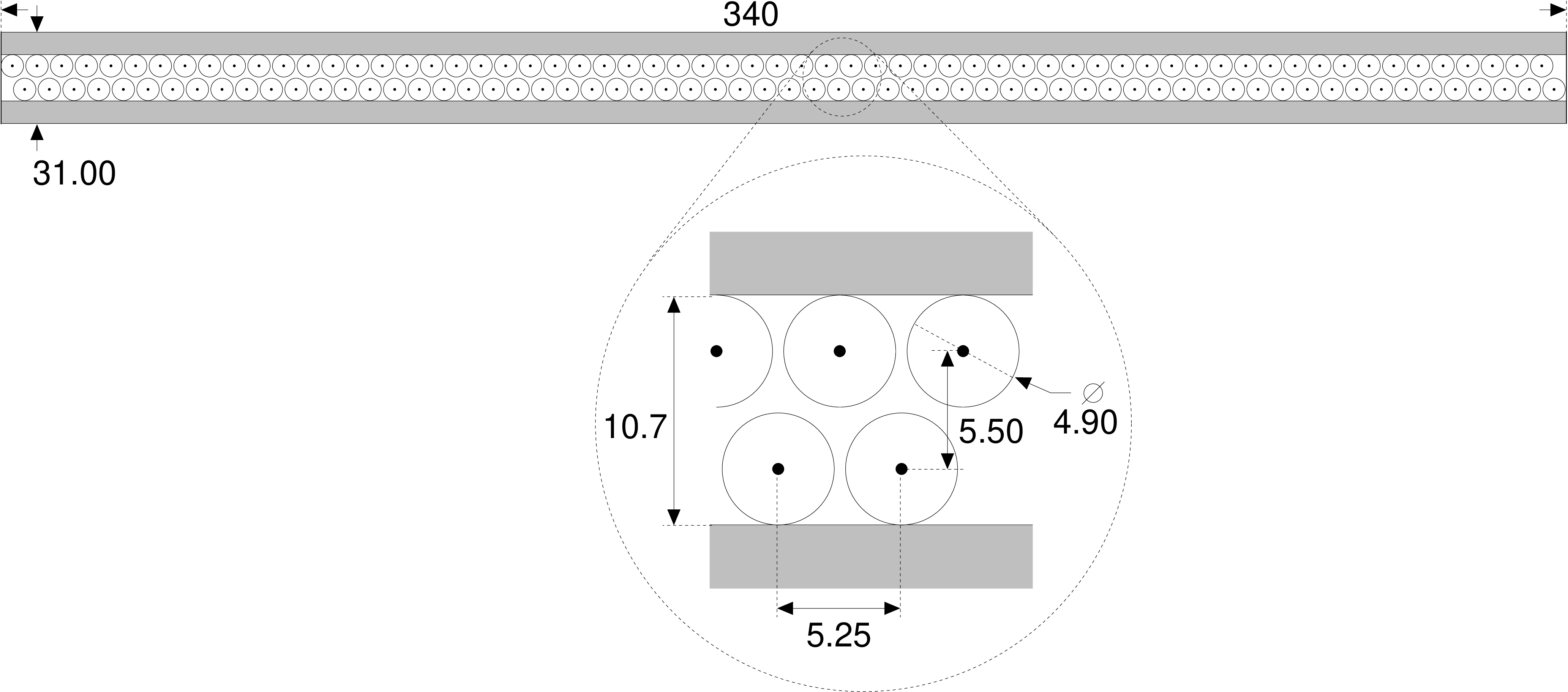}}
    \put(130,-5){\includegraphics[bb=0 0 456 298,scale=0.24]{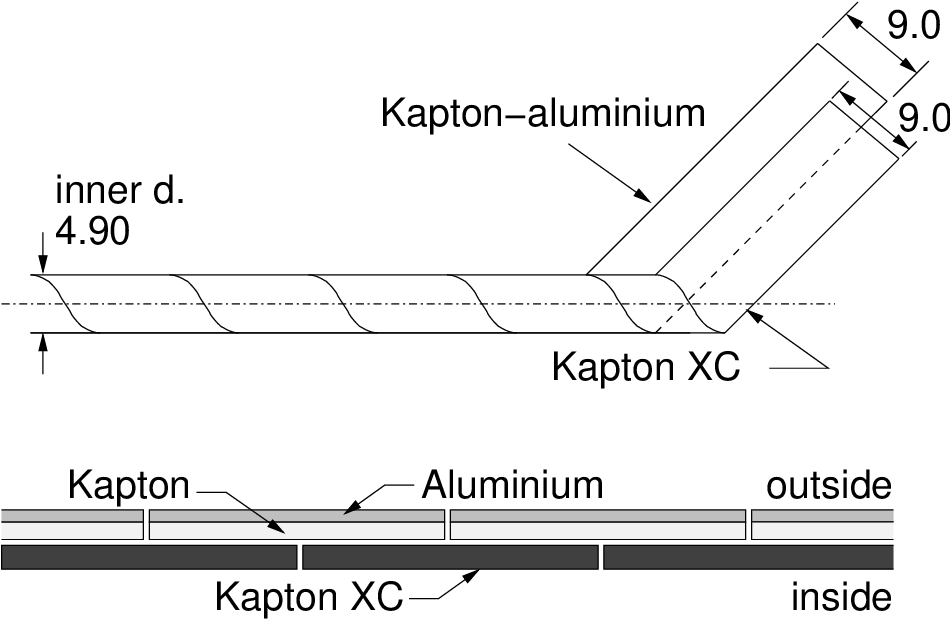}}
    \put( 20,50){{\em (a)}}
    \put(110,0){{\em (b)}}
    \end{picture}
    \caption[Outer Tracker module cross section]{\em (a) Module cross section. 
(b) The straws were winded using two foils, kapton-XC and a laminate of kapton and aluminium.}
    \label{fig:ot:modxsec}
\end{center}
\end{figure}

The straw tubes were 2.4~m long and 4.9~mm in diameter, and were filled with a gas
mixture of Ar(70\%)-CO$_2$(30\%). The anode wire was made of 25$\mu$m gold plated
tungsten wire, whereas the cathode consisted of an inner foil of 40~$\mu$m
electrically conducting carbon doped Kapton-XC and an outer foil of 25~$\mu$m
Kapton-XC with 12.5~$\mu$m aluminium.  The straws were glued to sandwich panels
with 120~$\mu$m carbon-fibre skins and a 10~mm Rohacell core. Finally, the
panels were joined by 400~$\mu$m thick carbon fibre sidewalls, resulting in a
standalone detector module. A sketch of the module layout is shown in
Fig.~\ref{fig:ot:modxsec}.  The panels and the side walls were covered by a
laminated foil of 25~$\mu$m Kapton and 12.5~$\mu$m aluminium to guarantee
gas-tightness of the box and to provide a closed Faraday cage. Spacers at the
ends of the module separated the two panels apart and pass the gas to the
module. All glueing steps were performed using Araldite AY103-1 with the hardener
HY991, cured at room temperature. To enhance the viscosity silica bubbles are
added.

The module production had started in April 2004 and extensive quality and performance 
measurements were performed. For example, signal propagation and analog electronics
characteristics were studied extensively with the use of a 2 mCi $^{90}$Sr source 
(with 1.7cm diameter collimator),  inducing a current of 3 nA.
In June 2005 it was discovered that, despite extensive aging tests in the R\&D
phase~\cite{ref:vienna04}, the module used for these tests suffered from gain loss 
at the same location as the irradiation.
These tests were not performed under well-defined conditions
and numerous external pollutants were considered.
In October 2005 a 20 mCi $^{90}$Sr source was used in a different setup
to study various gas mixtures,
(inducing about 100 nA per wire for about 12 hours)  and again gain loss was observed,
and a new systematic aging study campaign started.
In the meanwhile module production finished in November 2005.
This paper summarizes the findings, also described in Refs.~\cite{Bachmann:2010zz},
\cite{Tuning:2011zzb} and \cite{vanEijk:2012dx}.

\section{Aging Phenomenon}

\subsection{Irradiation and monitoring setup}
Irradiation tests were carried out on a small selection of final modules using a
2~mCi\sr source. The high voltage on the anode wires was set at 1600 V and the
gas flow (with a mixture of 70/30 Ar/CO$_2$) was 20~l/hr, corresponding to
approximately one volume exchange per hour.  The source was collimated by a hole
with a diameter of 6 mm at a distance of 5~mm from the module, resulting in an
irradiated area of approximately 6$\times$6~cm$^2$.
\begin{figure}[!h]
\begin{center}
    \begin{picture}(250,110)(0,0)
    \put(0,110){\includegraphics[bb=120 20 446 400,scale=0.3,angle=-90,clip=]{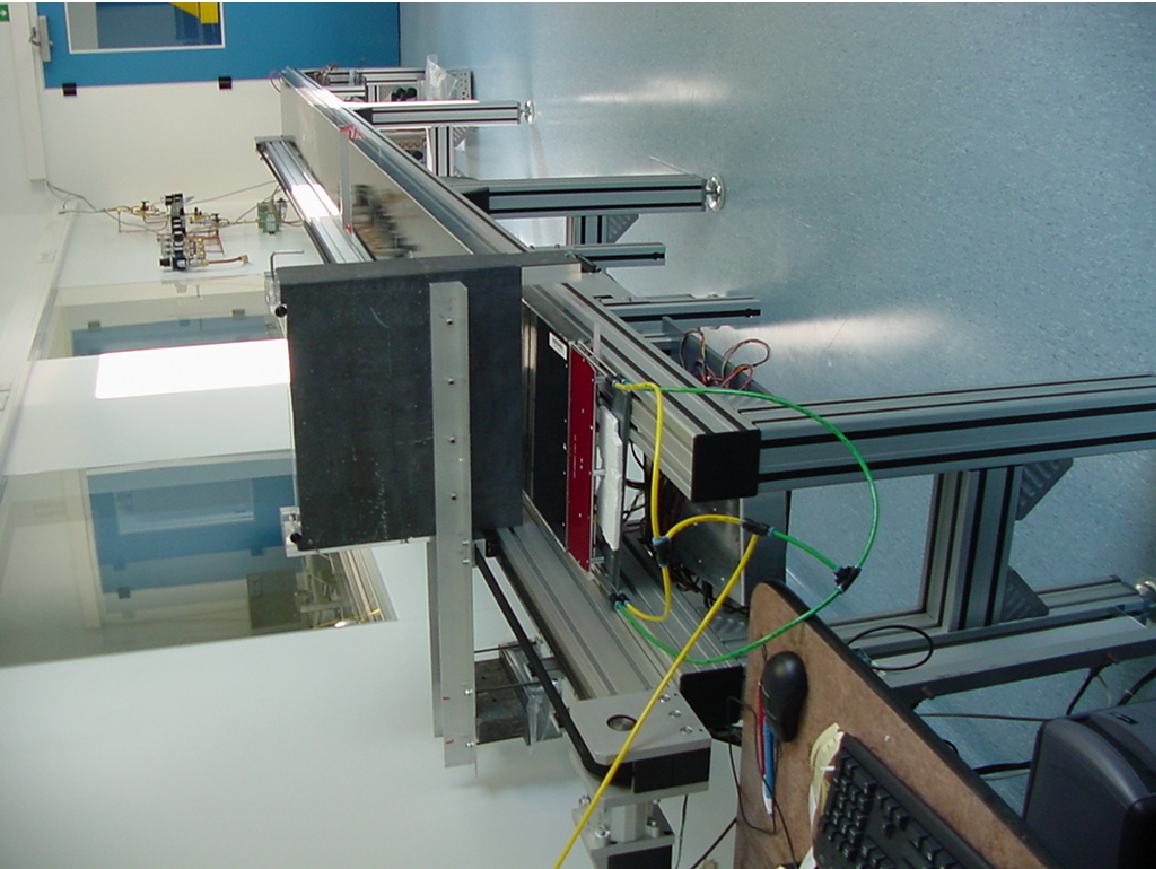}}
    \put(118,10){\includegraphics[bb=0 0 226 162,scale=0.62]{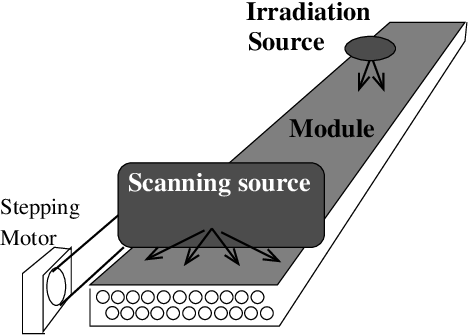}}
    \put(0,-5){{\em (a)}}
    \put(120,-5){{\em (b)}}
    \end{picture}
    \caption[Setup]{\em (a) Photograph of  the irradiation setup. The scanning source 
      with which the performance before and after irradiation is measured is also shown.
      (b) Schematic view of the irradiation and scanning setup.}
    \label{fig:setup}
\end{center}
\end{figure}

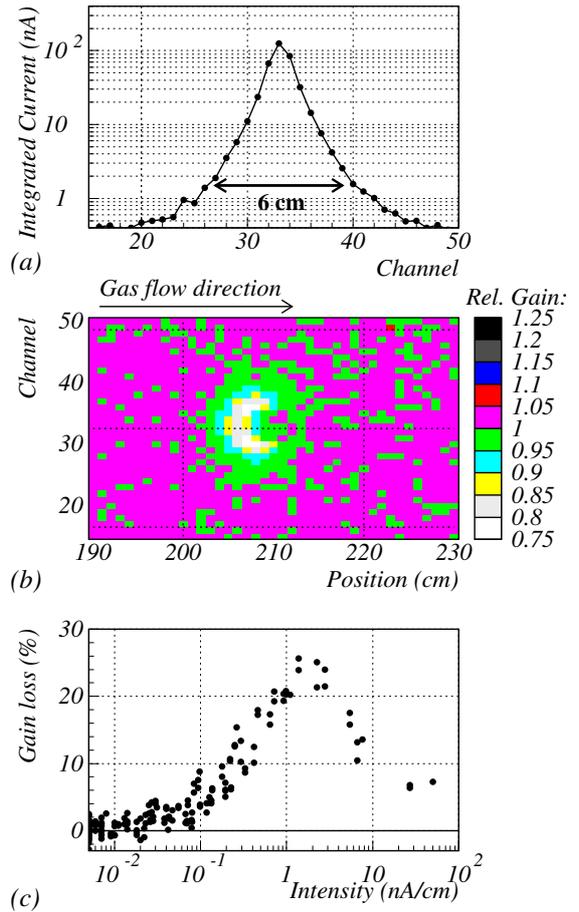
\begin{figure}[!t]
\begin{center}
    \begin{picture}(250,330)(0,0)
    \put(20,-5){\includegraphics[bb=15 20 300 520,scale=0.7]{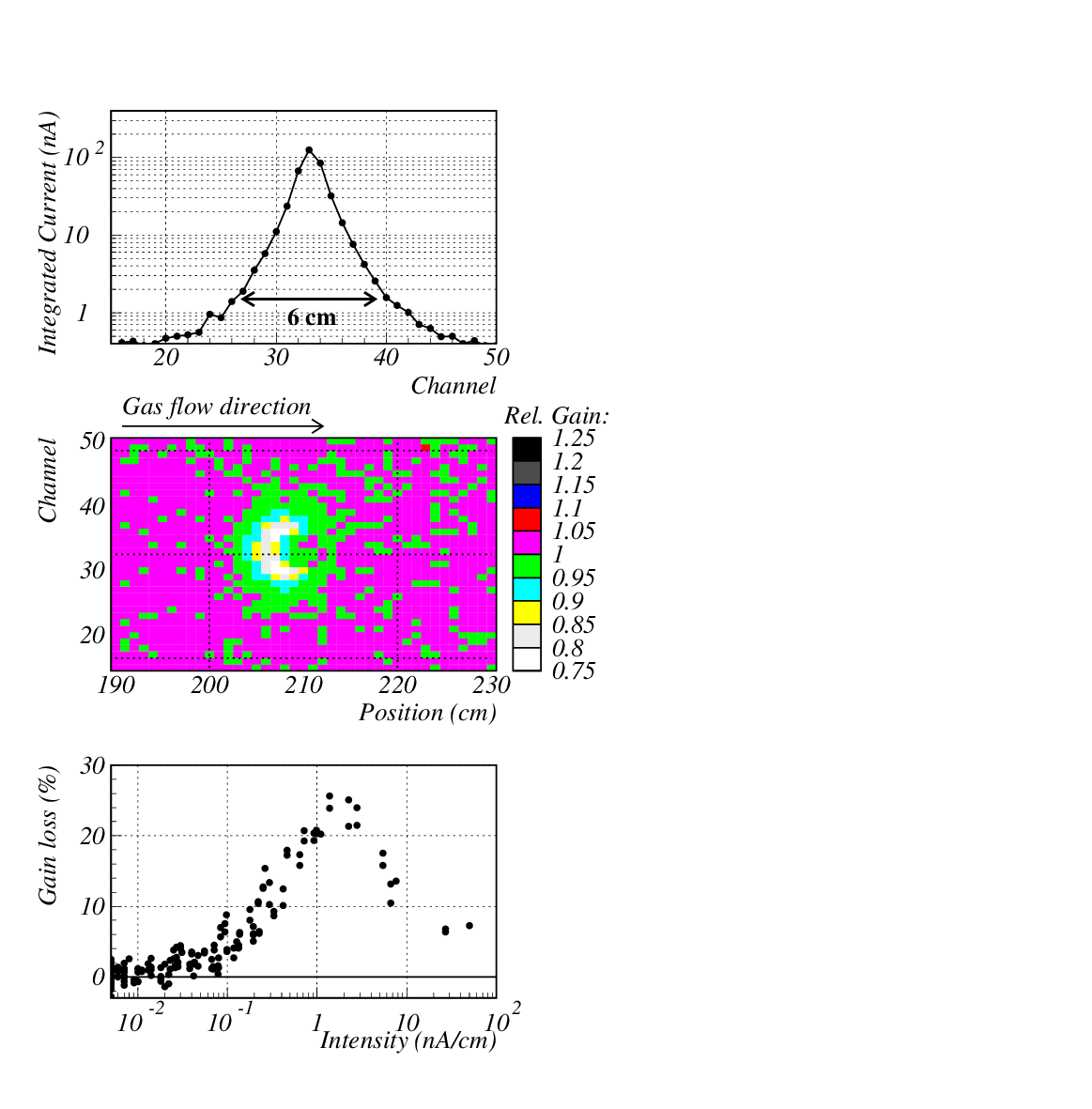}}
    \put(20,240){{\em (a)}}
    \put(20,120){{\em (b)}}
    \put(20,0){{\em (c)}}
    \end{picture}
    \caption[Source profile]{\em (a) The integrated current per wire during
    irradiation. (b) The ratio of two \sr scans before and after irradiation
    shows the relative gain loss after an irradiation of 20 hours. The source
    was centered on channel 32 on position 208cm.  (c) The gain loss is shown
    for each measurement (pixel of 0.5$\times$1~cm$^2$) as a function of the
    source intensity in that pixel. The gain loss was highest at moderate
    intensity, around 2~nA/cm.}
    \label{fig:prof-gain}
\end{center}
\end{figure}

Before and after irradiation the response of each wire in the module was checked
with a 20~mCi\sr source. The full module width was irradiated in steps of 1~cm
along the length and the corresponding wire current was measured and recorded.
The setup is depicted in Fig.~\ref{fig:setup}.
%
%
A typical example of the gain loss after an irradiation of 20 hours is shown in
Fig.~\ref{fig:prof-gain}b.  The gain loss was quantified by comparing the
2-dimensional current profile before and after irradiation, by means of dividing
the two current profiles.
The observed gain loss shows several distinguishing features:
\begin{itemize}
\item The gain loss was not proportional to the source intensity: directly under the
source the gain loss was less severe compared to the periphery. The gain loss for each 
measurement (corresponding to a pixel of $0.5\times 1$~cm$^2$) 
is shown as a function of the irradiation intensity
in Fig.~\ref{fig:prof-gain}c.
This dependency was unchanged when the module was irradiated 
at different values of the high voltage, or with different source strengths.
\item The gain loss occurred mainly upstream the source position,
and was worse for larger gas flow.
Presumably due to the creation of ozone in the avalanche region,
the gain loss was prevented downstream, see Section~\ref{sec:additives}.
\item The gain loss was large, upto 25\% for an integrated dose of 0.1~mC/cm
at an intensity of 2~nA/cm.
\end{itemize}

\subsection{Wire inspection and outgassing}
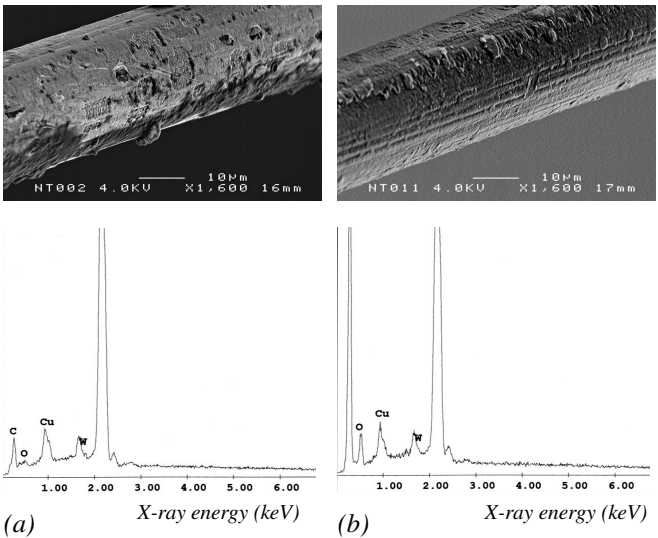
\begin{figure}[!h]
  \begin{center}
    \begin{picture}(350,200)(0,0)
      \put(  0,120){\includegraphics[bb=20 20 535 340,clip=,scale=0.23]{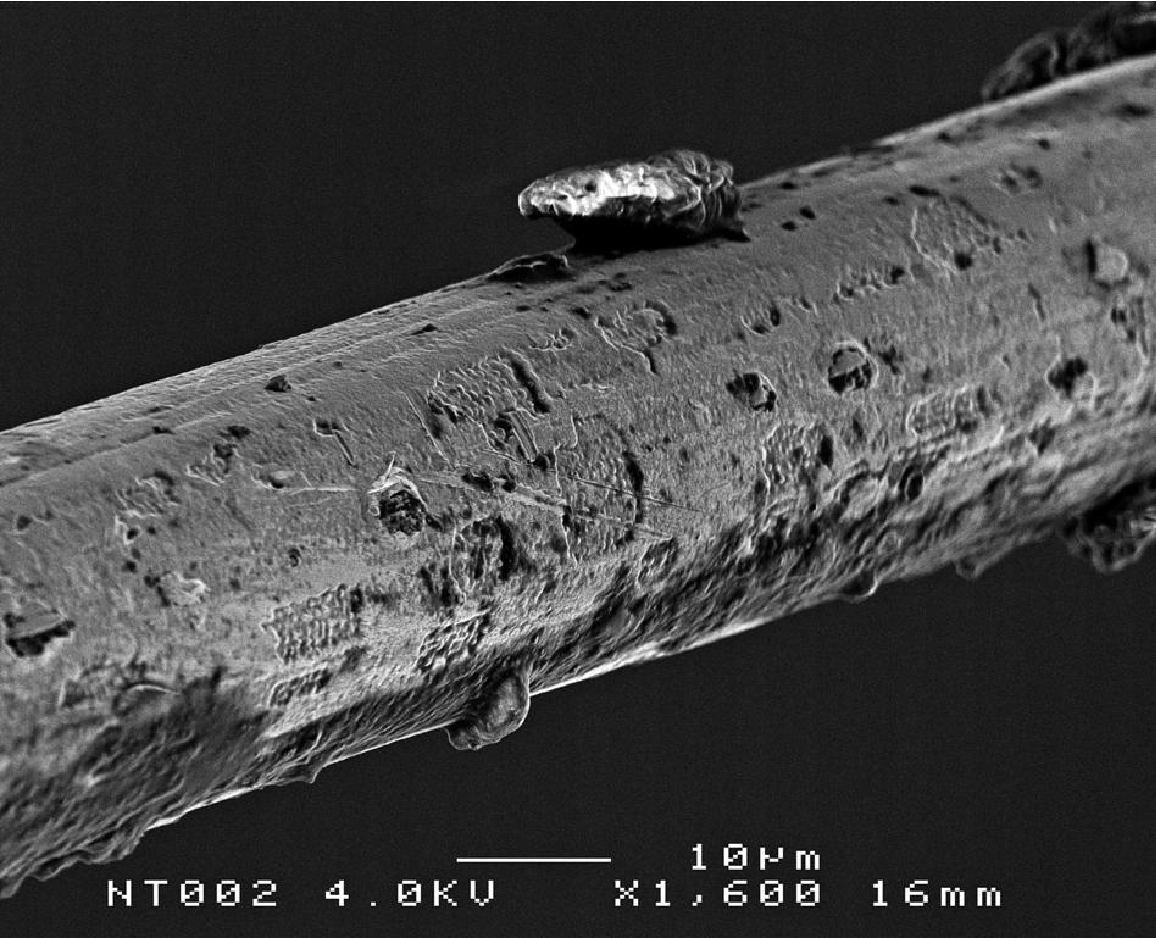}}
      \put(125,120){\includegraphics[bb=20 20 535 340,clip=,scale=0.23]{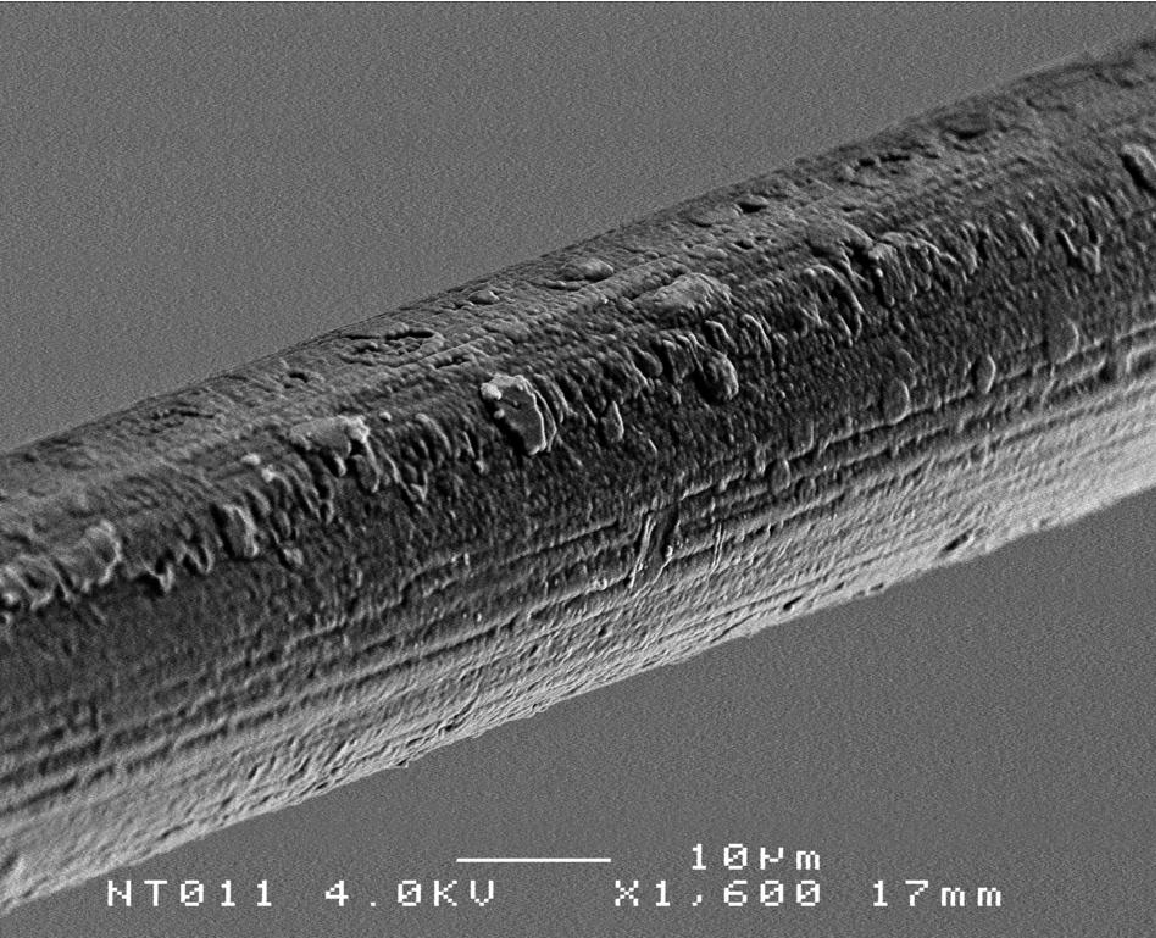}}
      \put(  0, 10){\includegraphics[bb=28 32 458 400,clip=,scale=0.27]{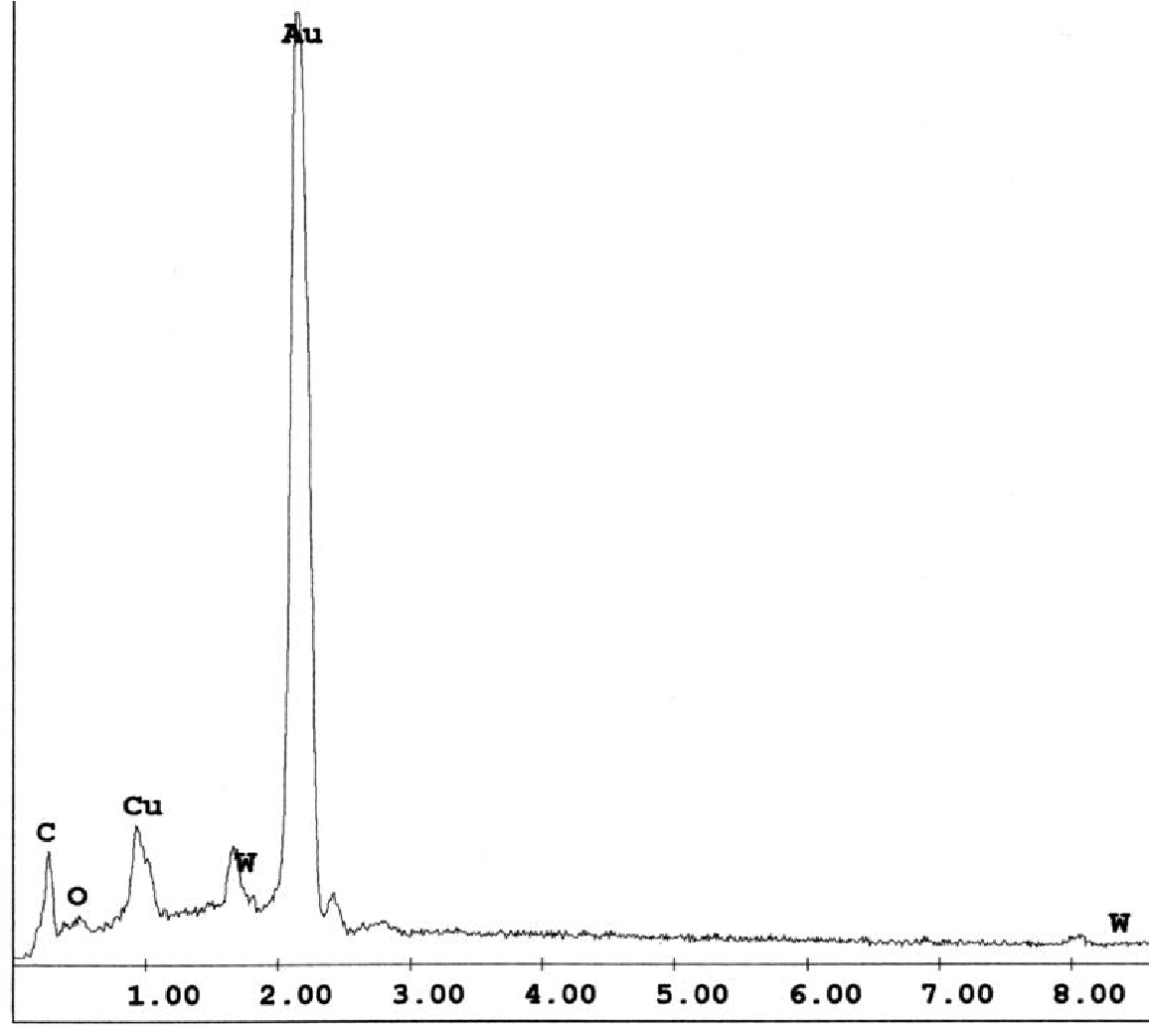}}
      \put(125, 10){\includegraphics[bb=30 29 460 398,clip=,scale=0.27]{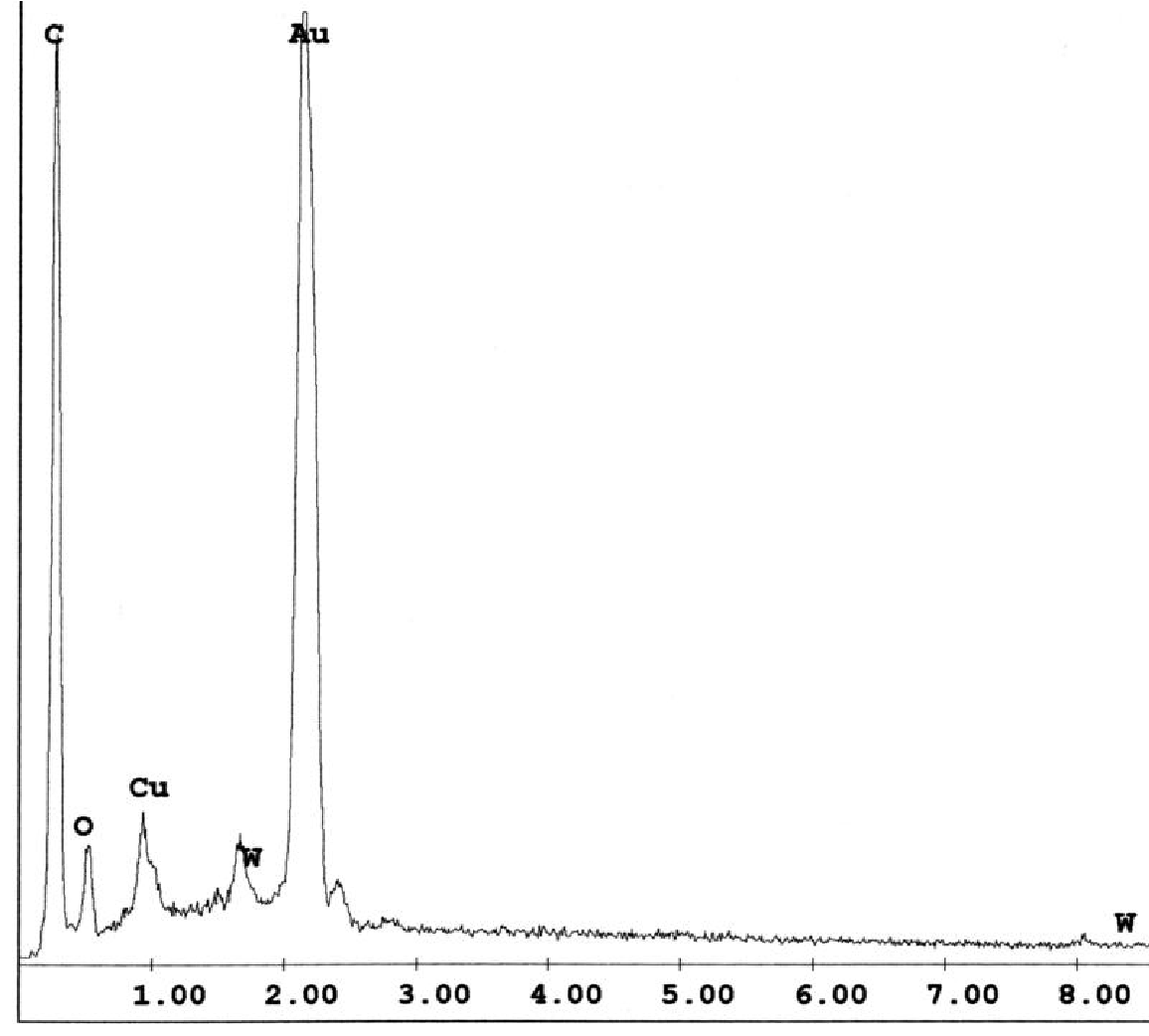}}
      \put(50,  0){{\em\footnotesize X-ray energy (keV)}}
      \put(180,  0){{\em\footnotesize X-ray energy (keV)}}
      \put(0,   -5){{\em (a)}}
      \put(125, -5){{\em (b)}}
    \end{picture}
    \caption[SEM picture of wire]{\em (a) A SEM picture and EDX spectrum is  shown for a
      sample of unirradiated outer tracker anode wire.
      (b) The same for an irradiated wire sample. A layer with a wax-like structure is observed,
      and a large amount of carbon is seen in the EDX-spectrum, indicating the presence of
      carbon-hydrates.\label{fig:wire}}
  \end{center}
\end{figure}

Samples of the anode wire were removed from an irradiated module for inspection
with a sampling electron microscope (SEM). An irradiated wire with observed gain loss 
as described in the previous section, was compared with an unirradiated wire.
An electrically insulating coating was found on the irradiated wire, see Fig.~\ref{fig:wire}.
The deposits were analyzed by means of energy-dispersive X-ray spectroscopy (EDX)
which revealed the presence of carbon and indirectly that of hydrogen.

\begin{figure}[!b]
\begin{center}
    \begin{picture}(250,220)(0,0)
    \put( 10,118){\includegraphics[bb=0 0 523 270,scale=0.42]{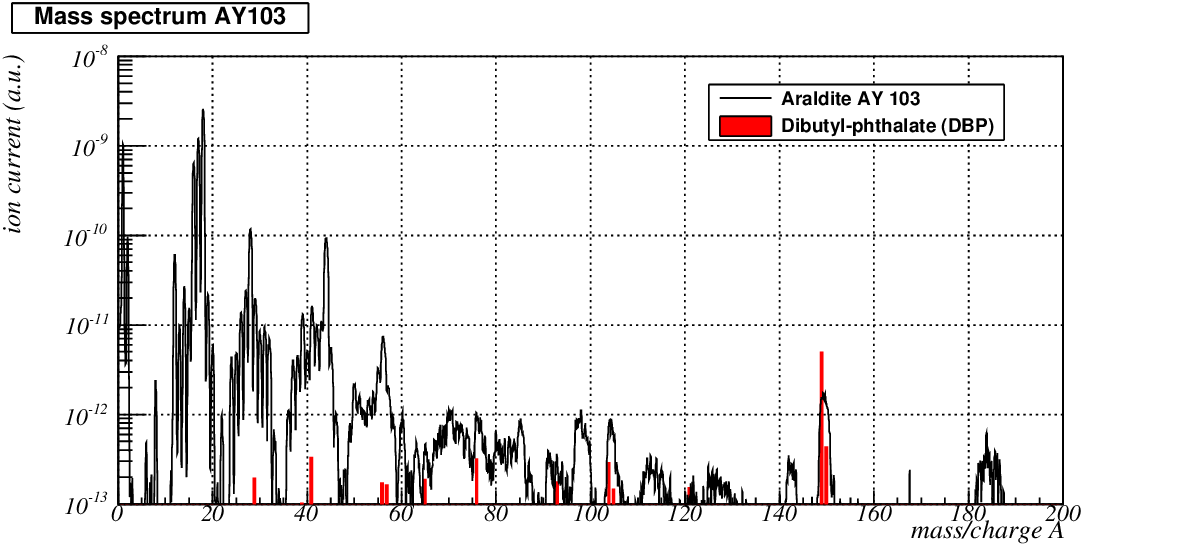}}
    \put(90,170){\includegraphics[bb=20 20 575 563,scale=0.09]{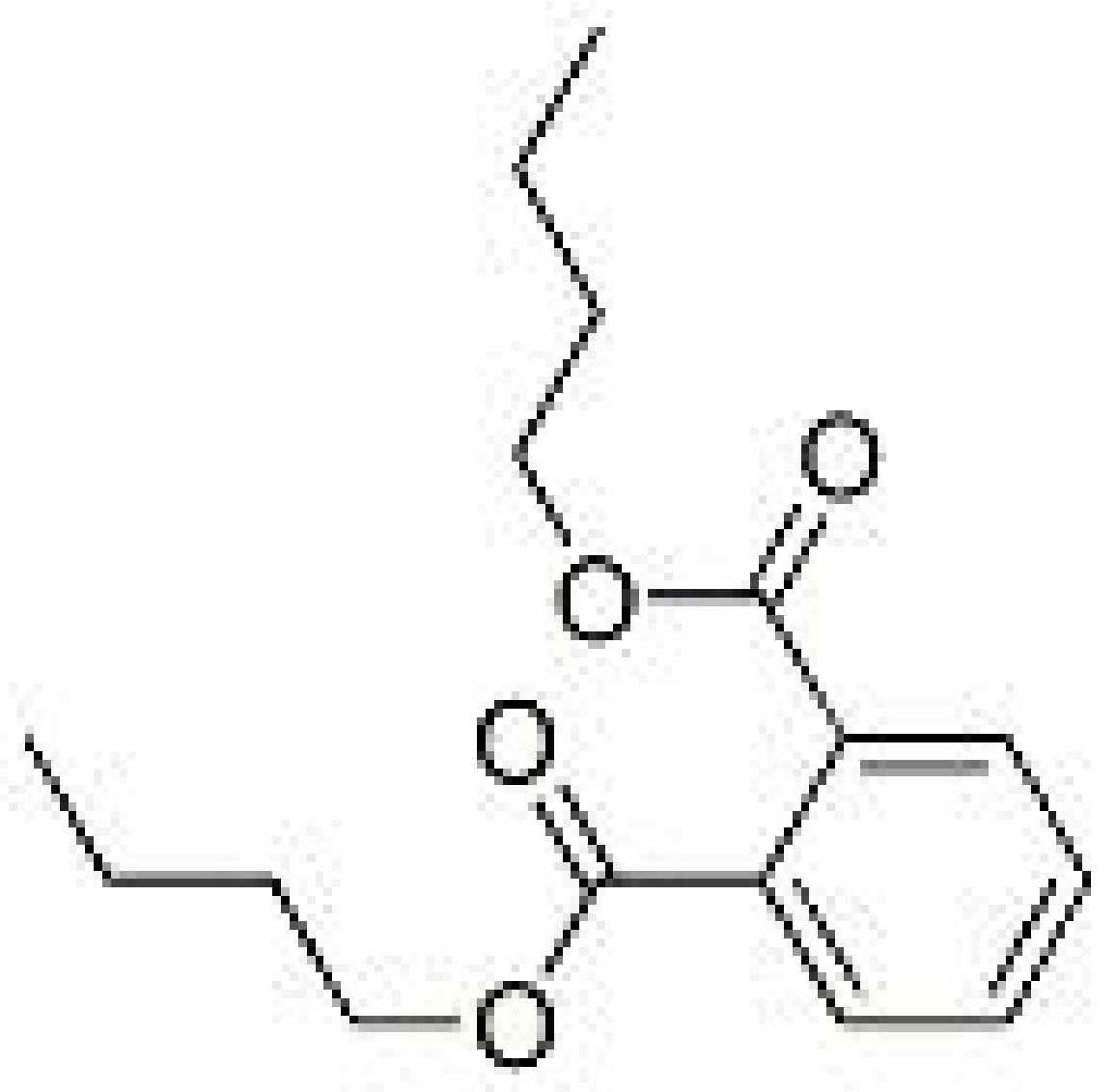}}    
    \put(  10, -5){\includegraphics[bb=0 0 523 270,scale=0.42]{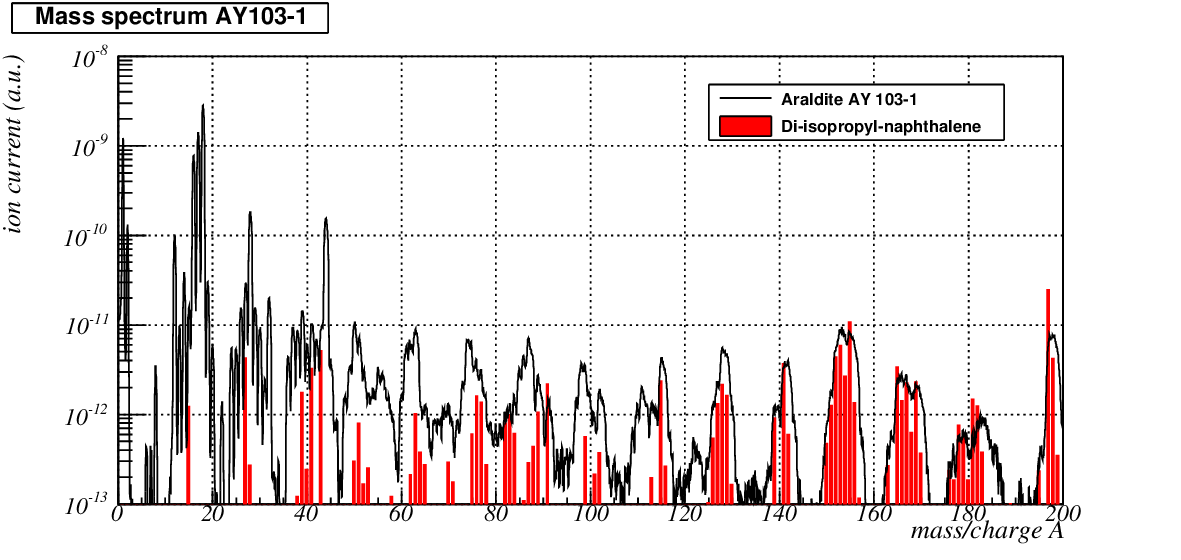}}
    \put(90, 60){\includegraphics[bb=20 20 575 409,scale=0.09]{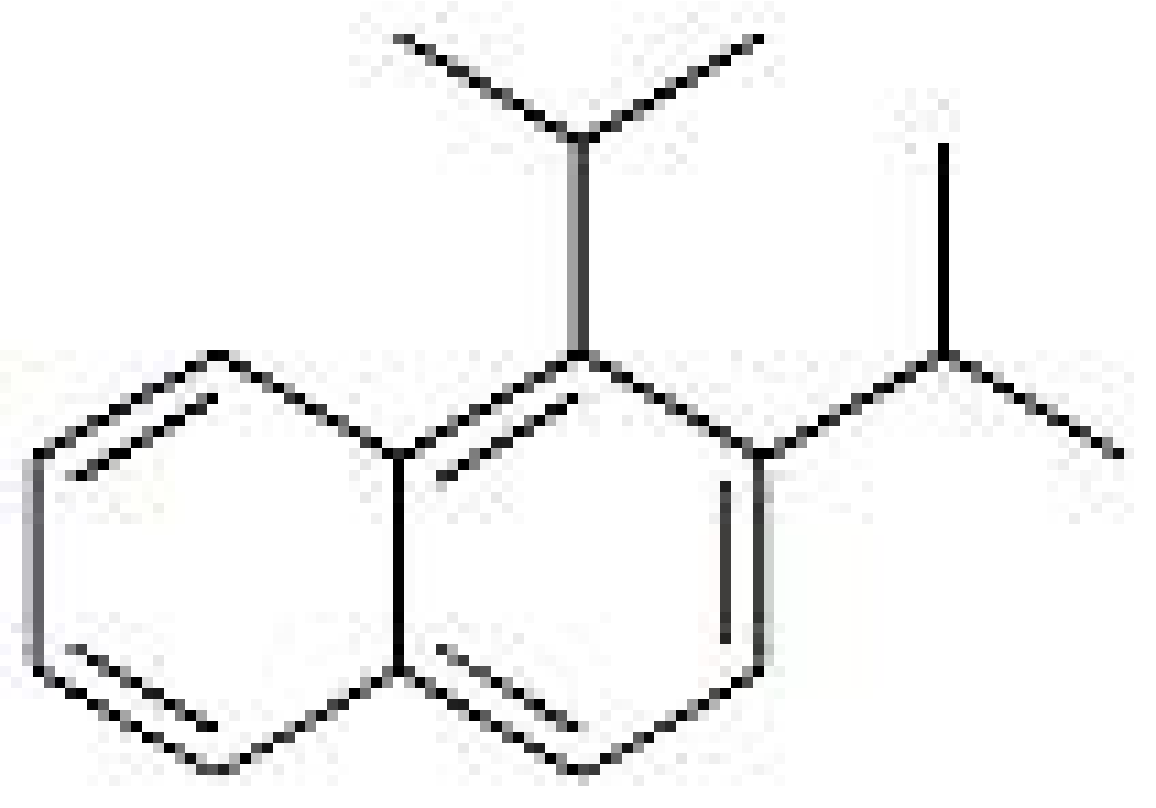}}    
    \put( 0,120){{\em (a)}}
    \put(60,210){C$_{16}$H$_{22}$O$_4$~:}
    \put( 0,  0){{\em (b)}}
    \put(70, 90){C$_{16}$H$_{20}$~:}
    \end{picture}
    \caption[Mass spectra]{\em (a) In the mass spectrum of AY103 traces of
      the plastifier dibutyl-phthalate (CAS nr.  84-74-2) are identified. (b) In the mass spectrum of AY103-1 traces of
      the plastifier di-isopropyl-naphthalene (CAS nr. 38640-62-9) are identified.
    The molecular structure of the plastifiers is also shown.}
    \label{fig:ms}
\end{center}
\end{figure}
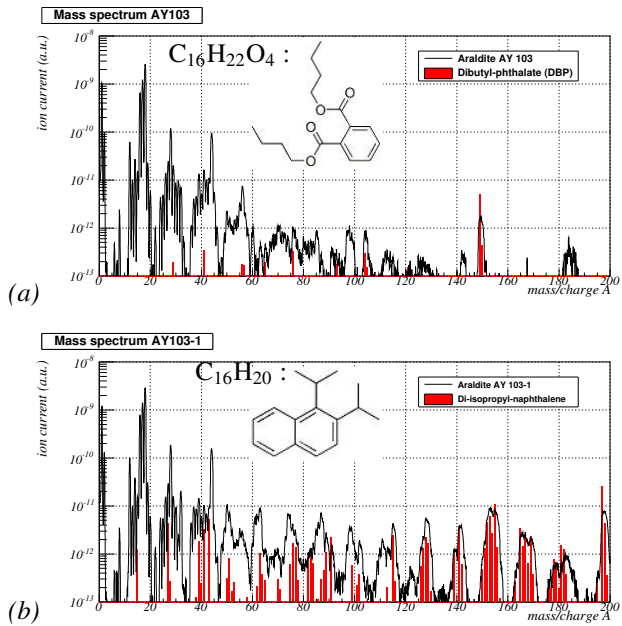

The outgassing of AY103 and AY103-1 was investigated by placing cured glue
samples in a vacuum chamber. The residual gas emitted by the glue was analyzed
with a quadrupole mass spectrometer. The quadrupole mass filter sorts the
produced ions according to their mass/charge ratio up to a value of 200 $u/e$.  The
resulting spectra are shown in Fig.~\ref{fig:ms}.  
The traces of DBP and di-isopropyl-naphthalene 
are identified in the mass spectra of AY103 and AY103-1, respectively.

\section{The Culprit}
A comprehensive review of materials in
gas detectors can be found elsewhere \cite{ref:ATL-glue}.
After extensive aging tests before 2004 the glue AY103 was identified for the 
module construction.
The manufacturer produced the last batch of AY103 with plastifier dibutyl phthalate (DBP) in 2003, 
before switching to AY103-1 with plastifier di-isopropyl-naphthalene.

To identify the origin of the insulating deposits an aluminum test module was
constructed with a minimum of components, containing the straw tubes, wires,
wire locators and feed-through PCB's only.  The module was sealed with a large
O-ring.  No signs of gain loss were observed after 480 hours of irradiation with
maximum intensity of 75 nA/cm, corresponding to an integrated dose of
approximately 0.13~C/cm.  

\begin{figure}[!b]
\begin{center}
    \begin{picture}(250,120)(0,0)
    \put(0,  10){\includegraphics[bb=70 20 420 392,scale=0.25,clip=]{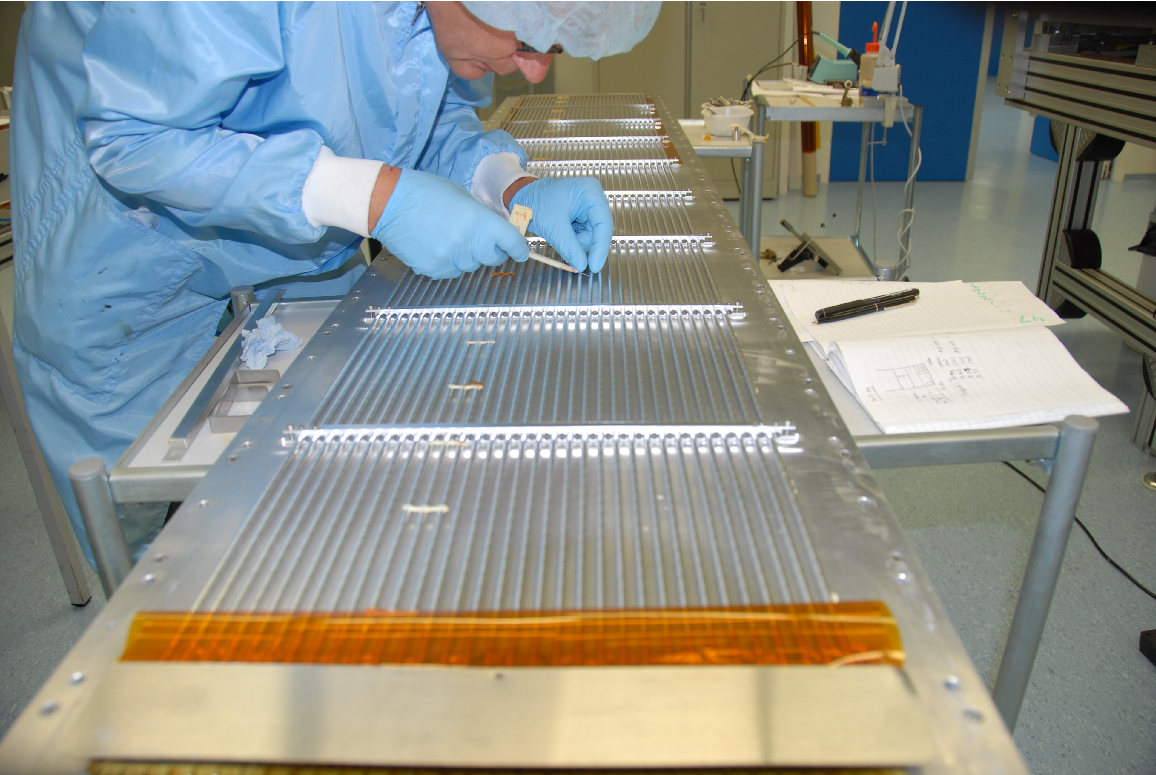}}    
    \put(0,-5){{\em (a)}}
    \put(100,-5){\includegraphics[bb=4 20 555 545,scale=0.27,clip=]{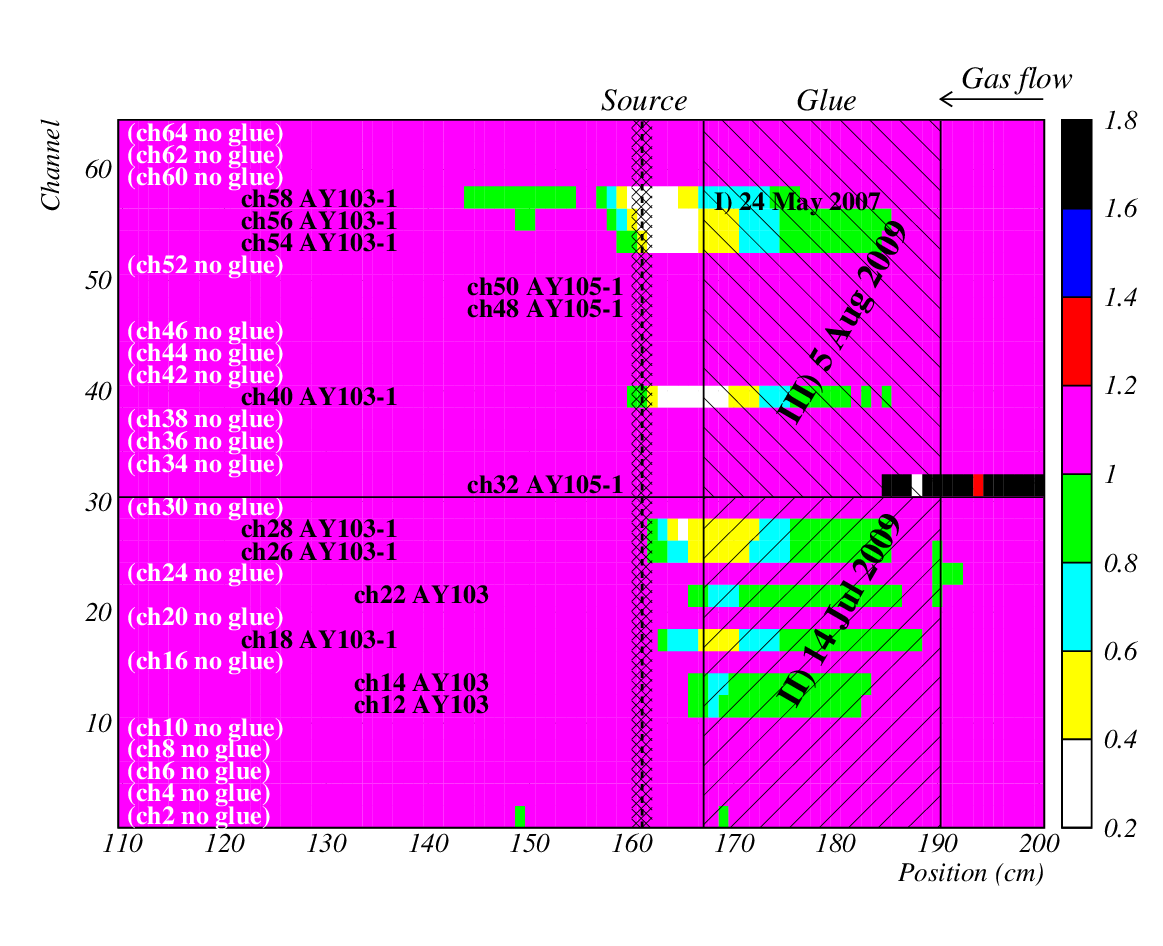}}
    \put(120,-5){{\em (b)}}
    \end{picture}
    \caption[Setup]{\em (a) Photograph of  the openable module during glue injection into a straw
      using a syringe. 
      (b)  The ratio of the response before and after the first 
      irradiation in August 2009 is shown
      for the full module. The type of injected glue is indicated
      per straw. The glue was injected between positions 167~cm and 190~cm
      on three different dates, as indicated in the figure. The source illuminated the
      module over the full width at position 161~cm.
      The gas flows from right to left.}
    \label{fig:aging-setup}
\end{center}
\end{figure}

In total seven straws were injected with AY103-1, three straws with AY103 and
three straws with AY105-1 (Fig.~\ref{fig:aging-setup}a).  The remaining 19 straws do not contain glue and are
used as reference.

Subsequently, the module was irradiated with a 20~mCi\sr source collimated such
that the source illuminates the module over the full width.  
Each straw was irradiated over a length of
approximately 4~cm, resulting in an irradiated area of approximately $34\times
4$~cm$^2$ .  The high voltage on the anode wires
was set to 1600 V and a gas mixture of Ar(70\%)-CO$_2$(30\%) was used. 
This corresponds to a gas amplification
of approximately $8 \times 10^4$. The gas flow was set to 20~l/hr, corresponding
to approximately one volume exchange per hour.  Before and after irradiation the
response of each wire in the module was measured with a scanning source.  
The full module width was
scanned in steps of 1~cm along the length,  and the corresponding wire current was measured and
recorded.

A strong reduction in the signal response due to the irradiation 
was observed in the channel containing
AY103-1, extending to a large area upstream the source position.  Also the
channel with AY103 showed a drop in the signal response upstream the source
position, albeit to a lesser extent compared to the epoxy containing
di-isopropyl-naphthalene, and limited to the region where the glue was inserted.
The channel containing the epoxy without any
plastifier, AY105-1, did not exhibit a gain drop after irradiation.

The summary of all results for the 32 straws irradiated in the first session
is given in Fig.~\ref{fig:aging-setup}b, that shows
the ratio of the response before and after irradiation.  
Different straws were injected with glue
at three different dates,  (1 week, 4 weeks and 2 years prior to irradiation, 
during which the module was not being flushed in the periods Jan-Apr 2008 and 
Mar-Jul 2009) 
but no dependence of the ageing rate with AY103-1 on these dates was observed, 
indicating that outgassing of the plastifier di-isopropyl-naphthalene was proceeding slowly.
The straws containing AY103-1 consistently showed a large drop in signal. 
Fast ageing effects have also been observed~\cite{ref:hydrocarbons} with other aromatic hydrocarbons,
like toluene and styrene with a molecular structure similar to that of naphthalene .

Finally, a full scale OT module was constructed with AY105-1. 
The module was irradiated over the full width at three different positions
for 500, 1060 and 980 hrs. Given the point of highest irradiation intensity,
this corresponds to an accumulated charge of 0.13, 0.28 and 0.25 C/cm, respectively.
A negligible gain loss of $5\% \pm 2\%$ was observed, indicating that this was not related
to the fast aging caused by AY103-1.

We concluded that outgassing of the plastifier in Araldite AY103-1 was the origin of the carbon
deposits on the anode wire, resulting in gain loss in the detector.

\section{Beneficial treatments}
The {\em shape} of the dependency of the gain loss on the irradiation intensity
remains unchanged when parameters of the irradiation tests are changed, such as:
high voltage, source type, source intensity, gas flow, gas mixture, humidity,
irradiation time, or flush time.  However, the maximum gain loss {\em does} vary,
depending on some of these parameters.  Beneficial effects on the maximum
gain loss are described in this section.

\subsection{Flushing}
\begin{figure}[!t]
  \begin{center}
    \begin{picture}(350,150)(0,0)
      \put(20,0){\includegraphics[bb=10 350 265 525,clip=,scale=0.75]{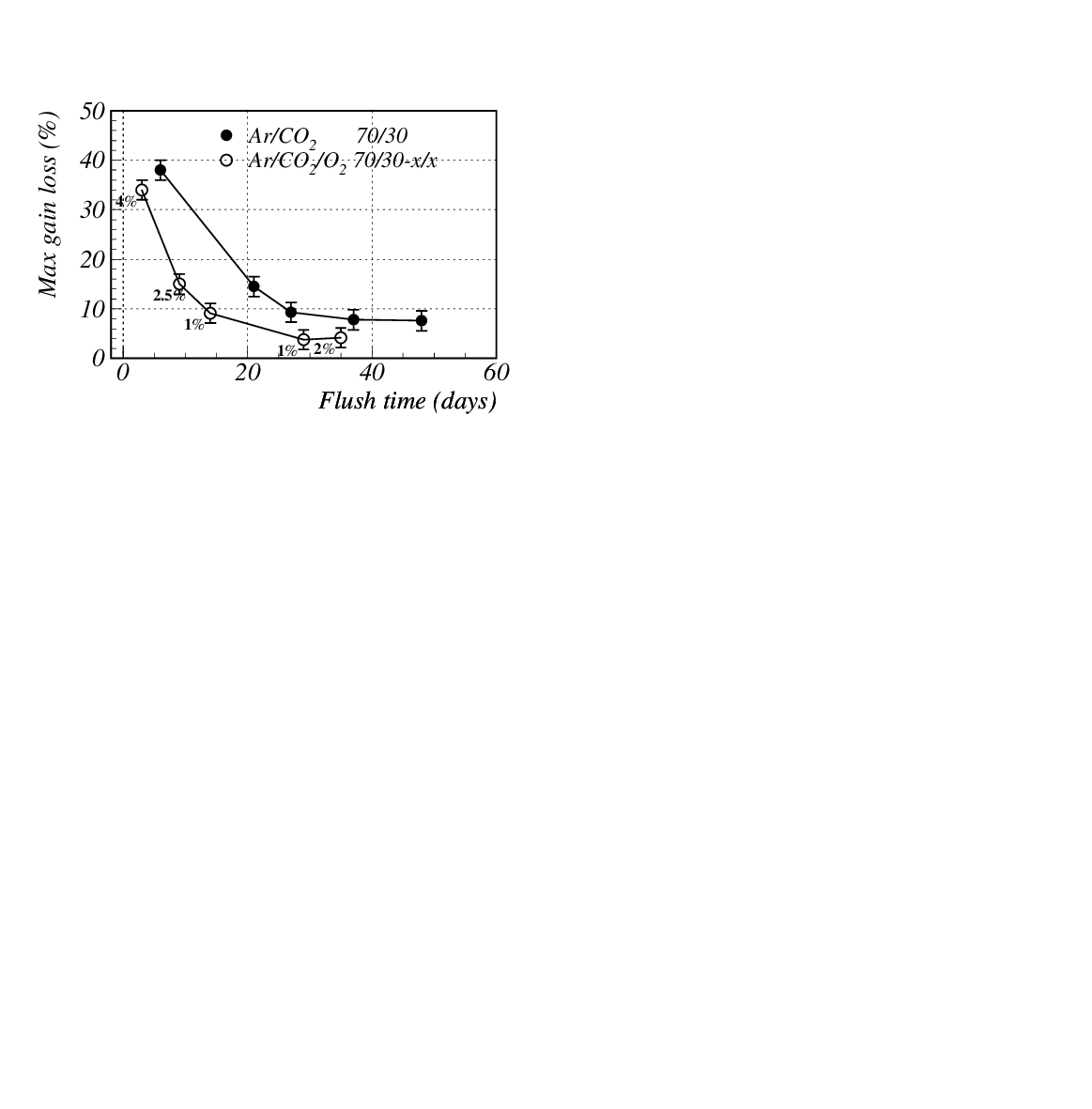}}
    \end{picture}
    \caption[Effect of Flushing]{\em The maximum gain loss decreased as a function
      of flushing time. Also, the aging rate was smaller when a few percent of oxygen is 
      added to the gas mixture, varying between 1\% and 4\%. 
      The improvement did not strongly depend on the exact amount 
      of oxygen.\label{fig:flush-oxy}}
  \end{center}
\end{figure}

Given the fact that the araldite AY103-1 glue used in construction was a
necessary ingredient to cause the gain loss in the OT detector, long term
flushing was expected to transport away the vapours originating from outgassing
of the glue.  Indeed, Fig.~\ref{fig:flush-oxy} shows the maximum gain loss
caused by an irradiation of 20hrs as a function of flush time. The aging rate
decreased significantly.  All OT modules had been flushed continuously since
the completion of installation in the LHCb experiment in Spring 2007.

In addition, experiments in the laboratory showed that heating the modules
at 40$^o$C might have accelerated the outgassing of the glue, although the effect on
the aging rate differed from module to module. All modules in the experiment
were heated for two weeks at 35$^o$C while flushed at 0.5 volume exchanges per
hour.

\begin{figure}[!t]
  \begin{center}
    \begin{picture}(350,230)(0,0)
      \put(30,0){\includegraphics[bb=15 180 265 520,clip=,scale=0.7]{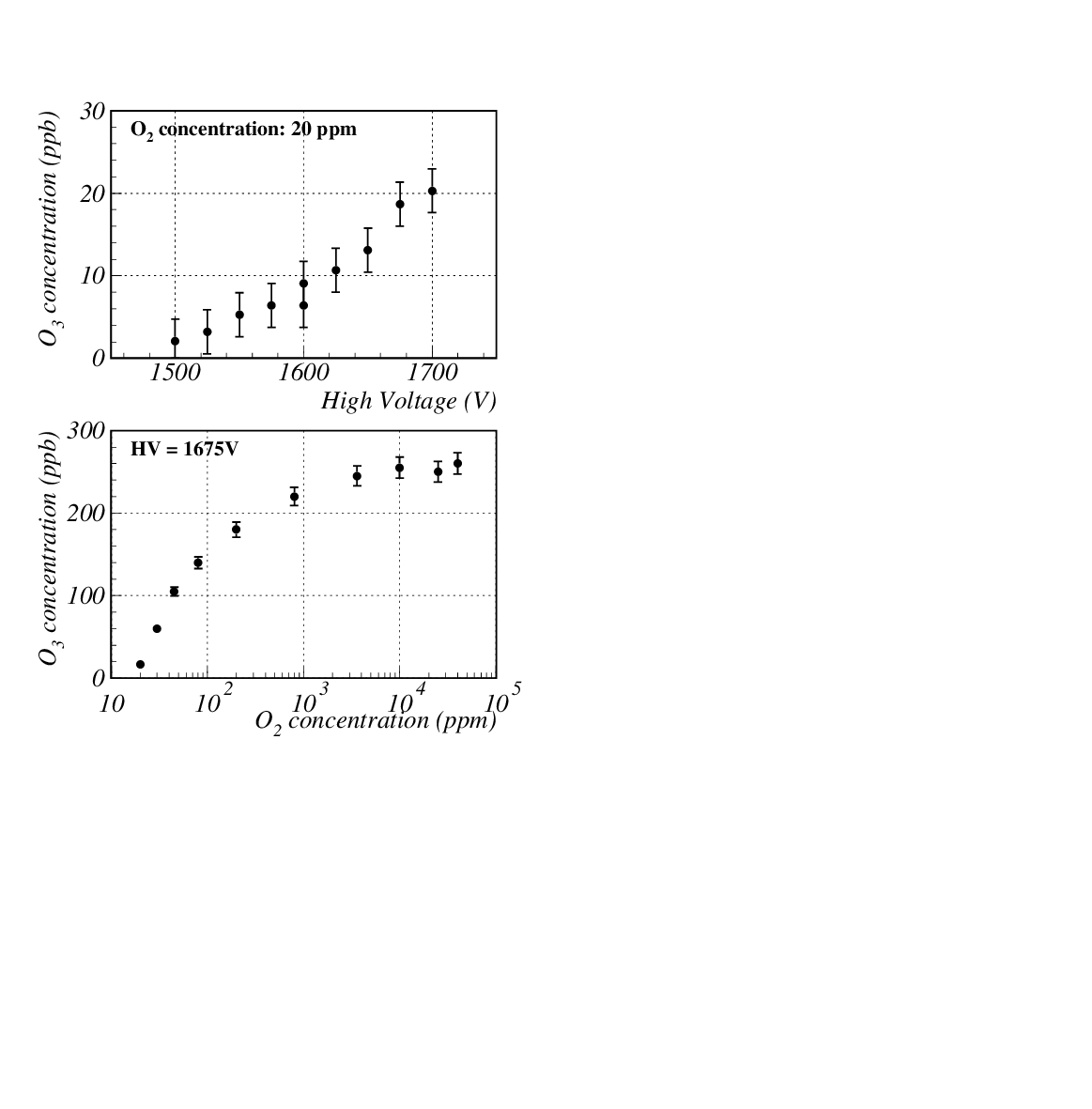}}
      \put(10,130){{\em (a)}}
      \put(10, 10){{\em (b)}}
    \end{picture}
    \caption[Ozone concentration]{\em (a) The ozone concentration increased for larger 
      values of the high voltage, i.e. for larger amplification, indicating that
      ozone was produced in the avalanche. (b) The ozone concentration as a function
      of oxygen concentration. Above 1\% O$_2$ the ozone concentration was constant.}
    \label{fig:ozone}
  \end{center}
\end{figure}

\subsection{Additives}
\label{sec:additives}

Oxygen has been used in other HEP experiments in the context of irradiation
damage and gas detectors~\cite{ref:ATL-O2,ref:CDF-O2}.  Tests with the OT show
that the aging rate for gas mixtures with O$_2$ is reduced by approximately a
factor two.  The improvement in aging rate with flushing time was similar for
the nominal gas mixture Ar/CO$_2$ 70/30, as compared to the gas mixture with a
few percent O$_2$ added, see Fig.~\ref{fig:flush-oxy}.  The amount of oxygen had
been varied between 1\% and 4\%, but the improvement did not strongly depend on
the exact amount of oxygen.

Due to the small diameter of the OT straw tubes, the average signal height at
the anode wire was only reduced by 11\% (29\%) when 2.5\% (4.5\%) O$_2$ was added
to the gas mixture.  Studies performed with the Garfield and Magboltz programs
indicated that the drift speed was not affected.

A sensitive ozone (O$_3$) meter had been placed at the gas outlet during
irradiation tests of OT modules. The ozone concentration increased with
increased high voltage (Fig.~\ref{fig:ozone}a), indicating that ozone was formed
in the avalanche region.  The production of ozone under the source was presumably
the reason that no gain loss is observed downstream of the source. This was
consistent with the observation that the aging rate was larger for increasing
gas flow, when the produced ozone was transported away more efficiently.

The concentration of O$_3$ had also been determined for various oxygen
concentrations, see Fig.~\ref{fig:ozone}b. For increasing oxygen concentration,
larger amount of ozone had been measured, which might explain the beneficial
effect of oxygen on the aging rate. Above 1\% of oxygen, the ozone
concentration did not increase, supporting the observation that no difference
in the aging rate was observed between 1\% and 4\% O$_2$.

\subsection{HV training}
Once the detector suffered from gain loss caused by the irradiation, the
insulating deposits could be removed by applying an high voltage at elevated
values.  Applying a reverse bias around -1450~V leads to fluctuating large dark
currents above 1~$\mu$A. After a period of 24 hrs the gain was recovered.
Similarly, a positive bias between 1850~V and 1920~V leads to dark currents of
about 10~$\mu$A per wire.  The gain was recovered in most cases after 20 hours.
The microscopic mechanism was unclear, and could have been related to plasma sputtering of the wire surface
or elevated temperatures at the location of the discharges.

The procedure with positive voltage had the advantage that the currents were more
stable, being less sensitive to trips in the power supply.  Secondly, since the
OT detector was operated with positive bias, power supplies for positive voltages
were readily available for the whole detector.

Potential damage to the anode wire had been investigated in the scanning
electron microscope after a treatment of 35 hrs at 1880~V, subject to a current
of 50$\mu$A.  No signs of mechanical damage to the gold layer were observed. In
addition, the relative peak heights of gold (Au) and tungsten (W) in the EDX
spectrum indicated that the gold layer was undamaged.

\begin{figure}[!t]
\begin{center}
    \begin{picture}(250,160)(0,0)
    \put(-10,0){\includegraphics[bb=10 420 600 810,scale=0.44]{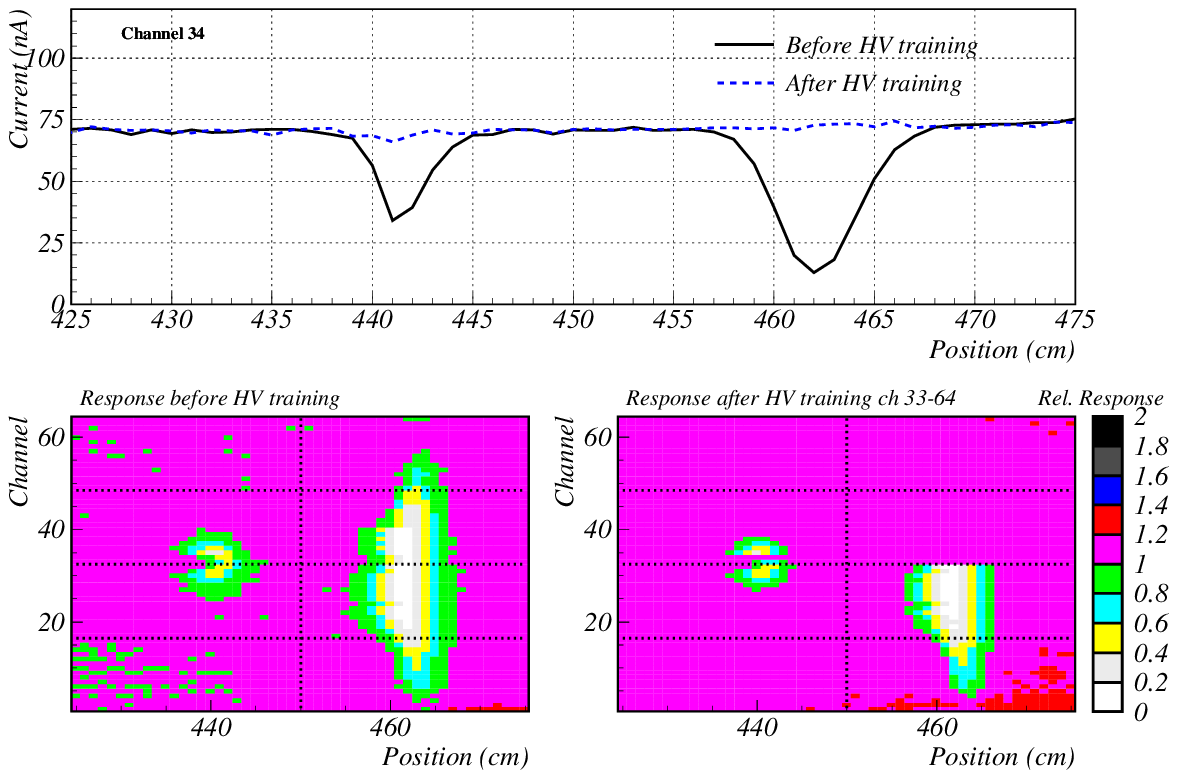}}
    \put(-10, 100){{\em (a)}}
    \put(-10,  0){{\em (b)}}
    \put(130,  0){{\em (c)}}
    \end{picture}
    \caption[HV training]{\em (a) The signal response is shown for one channel 
      before and after applying 
      a large positive high voltage of 1940~V. The signal decrease around 441~cm and 462~cm
      is due to gain loss induced by irradiation tests. (b) The signal response is shown
    for all 64 channels before the high-voltage treatment. 
    (c) The signal response is shown for all 64 channels after applying a high voltage of 1940~V
    to channels 33-64. The full signal gain is recovered in most places.} 
    \label{fig:hvtraining} 
\end{center}
\end{figure}

\section{Detector performance in LHCb}
The performance of the Outer Tracker detector in Run-1 and Run-2 of the LHC
has been according to (or better than) design specifications
\cite{LHCbOuterTrackerGroup:2013epe, LHCbOuterTrackerGroup:2017hwu}.
The single hit efficiency was measured to be above 99\%, while the noise rate was
below 40 kHz (which implies a 0.1\% occupancy of noise hits per event at 40 MHz readout rate).
Due to the low amplifier threshold setting, the effect of time-walk was minimized, and
a time resolution of 3~ns was obtained, leading to an intrinsic position resolution of
better than 180~$\mu$m.
The relatively large straw diameter of 4.9~mm would lead to unacceptably large
detector occupancies at the five-fold increased  instantaneous luminosity of 
$2\times 10^{33}$cm$^{-2}$s$^{-1}$ at Run-3 of the LHC, which necessitated the
replacement of the Outer Tracker by the Scintillating Fiber Tracker.

\section{Conclusion}
The LHCb Outer Tracker detector modules had shown to suffer from significant gain loss after
irradiation in the laboratory at moderate intensities, peaking at about 2 nA/cm.
During irradiation an insulating layer containing carbon is formed on the anode
wire.  The aging is caused by contamination of the counting gas due to outgassing
of the glue used in construction, namely araldite AY103-1.  The gain loss is
concentrated upstream the gas flow, due to the beneficial effect of ozone
produced under the source and transported downstream.  The aging rate was
reduced by longterm flushing and by the addition of a few percent of O$_2$ to
the gas mixture. Finally, the LHCb Outer Tracker has not shown any sign of gain degradation
over its lifetime within the LHCb detector between 2008 and 2018, presumably
because of the intrinsic beneficial effects of ozon production in the entire volume of the 
Outer Tracker, overcompensating any malicious formation of insulating deposits.
The cure of applying a large positive high voltage (beyond the
amplification regime), to remove any insulating deposits, was not necessary.
Given the excellent performance of the Outer Tracker until its last days of operation, 
potential re-use of the detector can be foreseen.

\bibliographystyle{ieeetr}

\begin{thebibliography}{999}

\bibitem{LHCbOuterTrackerGroup:2013epe}
R.~Arink \textit{et al.} [LHCb Outer Tracker Group],
Performance of the LHCb Outer Tracker,''
{\it {JINST \textbf{9} (2014) no.01, P01002}}
[arXiv:1311.3893 [physics.ins-det]].

\bibitem{ref:vienna04} S.Bachmann {\it{et al.}},  {\it{The straw tube technology for the 
LHCb outer tracking system}}, Nucl. Instrum. Meth.A535:171-174, 2004.

\bibitem{Bachmann:2010zz}
S.~Bachmann, \textit{et al.}
{\it {Ageing in the LHCb outer tracker: Phenomenon, culprit and effect of oxygen}},
Nucl. Instrum. Meth. A \textbf{617} (2010), 202-205.

\bibitem{Tuning:2011zzb}
N.~Tuning,  \textit{et al.}
{\it {Ageing in the LHCb outer tracker: Aromatic hydrocarbons and wire cleaning}},
Nucl. Instrum. Meth. A \textbf{656} (2011), 45-50

\bibitem{vanEijk:2012dx}
D.~van Eijk, \textit{et al.}
{\it {Radiation hardness of the LHCb Outer Tracker}},
Nucl. Instrum. Meth. A \textbf{685} (2012), 62-69

\bibitem{ref:ATL-glue} M.~Capeans, {\it{Aging and materials: 
Lessons for detectors and gas systems
}}, Nucl.Instrum.Meth.A515:73-88, 2003.

\bibitem{ref:hydrocarbons}  H.~Andersson {\it{et al.}}, 
{\it Aging of proportional counters with gas mixtures containing impurities of aromatic hydrocarbons},
IEEE Nucl. Sci. Conf. Rec. Vol.4, (2004) 2053.









\bibitem{LHCbOuterTrackerGroup:2017hwu}
P.~d'Argent \textit{et al.} [LHCb Outer Tracker Group],
{\it {Improved performance of the LHCb Outer Tracker in LHC Run 2}},
JINST \textbf{12} (2017) no.11, P11016
[arXiv:1708.00819 [physics.ins-det]].


\end{thebibliography}


\end{document}